\documentstyle[epsf,12pt]{article}
\def\bn{\bigskip\noindent}
\def\t{\theta}

\def\n{\nonumber}
\def\t{\theta}

\def\C{\mbox{Ch}}
\def\T{\mbox{Tr}}
\def\o{\over}
\begin{document}
\begin{titlepage}
\renewcommand{\thefootnote}{\fnsymbol{footnote}}
 \font\csc=cmcsc10 scaled\magstep1
 {\baselineskip=14pt
 \rightline{
 \vbox{\hbox{hep-th/0010002}
       \hbox{UT-908}
       \hbox{September 2000}
       }}}
\bn
\begin{center}
 \huge{Singular Calabi-Yau Manifolds\\
and ADE Classification of CFTs
}   
\vspace{5ex}\\
\normalsize{Michihiro Naka
\footnote{email: naka@hep-th.phys.s.u-tokyo.ac.jp}
and Masatoshi Nozaki
\footnote{email: nozaki@hep-th.phys.s.u-tokyo.ac.jp}}
\vspace{5ex}\\
\normalsize{\it Department of Physics,\\
Faculty of Science, University of Tokyo,\\
Hongo 7-3-1, Bunkyo-ku, Tokyo 113-0033, Japan}
\end{center}
\vspace{5ex}
\abstract{
We study superstring propagations on the Calabi-Yau manifold 
which develops an isolated ADE singularity.
This theory has been conjectured to have a holographic dual description
in terms of $N=2$ Landau-Ginzburg theory and Liouville theory.  
If the Landau-Ginzburg description precisely reflects the information 
of ADE singularity,  
the Landau-Ginzburg model of $D_4,E_6,E_8$ and 
Gepner model of $A_2\otimes A_2, A_2\otimes A_3, A_2\otimes A_4$
should give the same result.
We compute the elements of $D_4,E_6,E_8$ modular invariants
for the singular Calabi-Yau compactification
in terms of the spectral flow invariant orbits of the tensor product theories
with the theta function which encodes the momentum mode of the Liouville 
theory.
Furthermore we find the interesting identity among characters in
 minimal models at different levels.
We give the complete proof for the identity.
}

\end{titlepage}
\section{Introduction}
\hspace{5mm}
Currently the study of superstring theory on singular Calabi-Yau manifolds
is active.
The important feature of superstrings propagating near singularities is the 
appearance of light solitons coming from the D-branes wrapped around the 
vanishing cycles.
This is the non-perturbative quantum effect in string theory 
in the sense that 
even after taking $g_s \to 0$, the VEV of dilaton
will blow up at the singular point.
In order to appear such an effect, the vanishing worldsheet theta angle
is necessary \cite{a}, which 
seems to make worldsheet CFTs singular \cite{w1}.
In contrast to the well-established 
perturbative description of smooth Calabi-Yau manifolds
like Gepner models \cite{g},
the worldsheet description of singular Calabi-Yau manifolds remains
to be investigated.
This situation is expected to bring us the new source 
of insights of stringy dynamics
related to the space-time singularity. 
Moreover, this set up only depends on the type
of the singularity, 
and has an interesting physical application that
the decoupled theory in this background
corresponds to, for example, four-dimensional CFTs
classified by ADE \cite{ad}.

Such a CFT can be described by the $N=2$ Landau-Ginzburg model with a
superpotential including a negative power of some chiral superfield
in order to push up the central charge to the right value for Calabi-Yau
manifolds,
and this peculiar term was handled by Kazama-Suzuki model \cite{kas}
for the non-compact coset $SL(2,{\bf R})/U(1)$ \cite{gv, ov}.
Recently, after the renowned AdS/CFT correspondence \cite{ma}, 
the approach based on a holographic point of view was proposed \cite{abks}.
In this approach, the sector of Landau-Ginzburg theory with a negative
power superpotential is replaced by Liouville theory \cite{gkp}.
This Liouville field corresponds to an extra
non-compact direction, which indicates holography.

For this description, the first consistency check 
is to make the modular invariant partition function on a torus, and to
see that the partition function vanishes.
This issue was systematically pursued in \cite{ov, es} for the singular
Calabi-Yau manifold with an isolated ADE singularity (or conifold \cite{mi}).
Furthermore, the extension to more complicated singularities was made \cite{y}.
In all the case, the 
string theory on the Calabi-Yau manifold with the ADE singularity
is treated as
$N=2$ minimal model classified by ADE and Liouville theory.

Now, we wish to pose the point of view taken throughout this paper.
In the Landau-Ginzburg description of $N=2$ minimal models, 
the following relation should hold \cite{mvw}:
\begin{equation}\label{rel}
D_4=A_2 \otimes A_2 ,\qquad E_6=A_2 \otimes A_3,\qquad E_8=A_2 \otimes A_4,
\end{equation}
with the appropriate projection in the CFTs.
This is because $D_4,E_6,E_8$ simple singularities are defined by
$x^3+xy^2=0,\; x^3+y^4=0,\; x^3+y^5=0$, up to quadratic terms.
We can use Gepner's construction \cite{g} 
for the tensor product theory in order to reproduce 
the feature of block diagonal $D_4, E_6, E_8$ modular invariants.
At first sight, one may think that $D_4$ invariant is not written by the
tensor product theory due to the term $xy^2$ in the above polynomial.
However we can rewrite the polynomial into the form $x^3+y^3$
by the suitable linear transformation of variables.
Above relation (\ref{rel}) 
can be reduced to the identity among characters in minimal 
models at different levels.
In principle, this can be predicted by the comparison
of weight and $U(1)$ charge of the irreducible representations in
minimal models.
But the physical meaning has been lacking.
The partition function 
for singular Calabi-Yau manifolds
is the slight generalization of well-known ADE
classification of modular invariant (for the recent discussion, see \cite{zu}).
If this partition function reflects the singularity of spacetime,
there shold exist
the reformulation of $D_4, E_6, E_8$ invariants by
the tensor product theory as in (\ref{rel})
and the identity among characters in minimal models.
This is our interpretation of the conventional simple singularity.
However, we cannot proceed the other non-diagonal
modular invariants $D_{n>4}, E_7$.

In order to make modular invariant of the tensor product theory, we use the
so-called spectral flow method \cite{eoty} which makes 
the block diagonalization and
the spacetime supersymmetry manifest. 
One reason for using the spectral flow method is 
that for singular K3 surface,
we can reproduce the
block diagonal elements of $D_4, E_6, E_8$ modular invariants
as the spectral flow invariant orbits by the tensor product theory 
of suitable minimal models. 
This cannot be seen only with the minimal models.
Due to the identity among the characters of minimal
models, the extension to the singular Calabi-Yau 3 or 4-folds is
straightforward.
Finally we give the proof of the proposed identity.

The paper is organized as follows. 
In section 2, we briefly review the construction of 
the modular invariant partition function 
in the system of superstrings on Calabi-Yau manifolds with the isolated ADE
singularity, based on the Liouville system and minimal model classified by ADE.
Section 3 is devoted to introduce the necessary background and notation
about the spectral flow method
used in the remainder of the paper.
In section 4, we propose our main results about the $D_4,E_6,E_8$
modular invariants.
We explicitly identify the
block diagonal elements of $D_4,E_6, E_8$ invariants by the spectral flow 
invariant orbits of Gepner models of $A_2\otimes A_2, A_2\otimes A_3,
A_2\otimes A_4$ and derive the identity among characters in minimal models 
at different levels.
Section 5 includes the conclusion and discussion.
In the Appendix A, we collect the formula
about the theta functions and characters of minimal models,
used in this paper extensively.
Appendix B includes the exact proof for the identity
among the characters in minimal models.

\section{Supertrings on singular Calabi-Yau manifolds}
\hspace{5mm}
In this section, we briefly review the construction of the 
modular invariant partition
function of non-critical superstring theory which is conjectured to
give the dual description of the Calabi-Yau manifolds 
with ADE singularity in the decoupling limit \cite{ov, es}.

\subsection{CFT for non-critical superstring}
\hspace{5mm}
Let us consider Type II string theory on the background
${\bf R}^{d-1,1}\times X_n$,
where $X_n$ is a Calabi-Yau $n$-fold ($2n+d=10$) with an isolated
singularity, locally defined by
$F(x_1,\dots,x_{n+1})=0$ in appropriate weighted projective space.
In particular, we concentrate on the isolated rational ADE singularity.
For singular 2-fold, or K3 surface, the following polynomials
define the singular geometries
\begin{eqnarray}\label{ade}
F_{A_{N-1}}\;&=&x_1^{N}+x_2^2+x_3^2,\qquad\;\;\quad(N\geq2)\nonumber\\
F_{D_{\frac{N}{2}+1}}&=&x_1^{N/2}+x_1x_2^2+x_3^2,\qquad(N:{\rm even}\geq6)
\nonumber\\
F_{E_6}\;\;\;\;\;&=&x_1^4+x_2^3+x_3^2,\\
F_{E_7}\;\;\;\;\;&=&x_1^3x_2+x_2^3+x_3^2,\nonumber\\
F_{E_8}\;\;\;\;\;&=&x_1^5+x_2^3+x_3^2.\nonumber
\end{eqnarray}
For singular Calabi-Yau 3,4-folds, we add $x_4^2, x_4^2+x_5^2$ to above
polynomials.
The quadratic terms do not change the type of singularity.

In the decoupling limit $g_s\to 0$, we obtain a non-gravitational,
and maybe non-trivial quantum theory on ${\bf R}^{d-1,1}$.
These $d$-dimensional quantum theories are expected to flow into 
non-trivial conformal RG fixed points in the IR limit.

According to the holographic duality \cite{gkp}, we have the dual description
of the above system in terms of non-critical superstrings on 
\begin{equation}
{\bf R}^{d-1,1} \times \left({\bf R}_{\phi}\times S^1\right)\times LG(W=F),
\end{equation}
where $LG(W=F)$ denotes the $N=2$ Landau-Ginzburg model with a
superpotential $W=F$.
And ${\bf R}_{\phi}$ denotes a linear dilaton background with the
background charge $Q(>0)$.
The part ${\bf R}_{\phi}\times S^1$ is described by the $N=2$ Liouville
theory \cite{kus1} 
whose matter content consists of bosonic fields $\phi,Y$
which parameterize ${\bf R}_{\phi}, S^1$, respectively, 
and their fermionic partners
$\psi^+,\psi^-$.
Then the $N=2$ superconformal currents are written as
\begin{eqnarray}
&&T\;\:\;=-\frac{1}{2}\left(\partial Y\right)^2-\frac{1}{2}
\left(\partial\phi\right)^2-\frac{Q}{2}\:\partial^2\phi-\frac{1}{2}
\left(\psi^+\partial\psi^+-\partial\psi^+\psi^-\right),\nonumber\\
&&G^{\pm}=-\frac{1}{\sqrt{2}}\psi^{\pm}\left(i\partial Y\pm\partial\phi\right)
\mp\frac{Q}{\sqrt{2}}\:\partial\psi^{\pm},\\
&&J\;\:\;=\psi^+\psi^--Q\: i\partial Y\nonumber,\label{sca}
\end{eqnarray}
which generate the $N=2$ superconformal algebra with central charge 
$c=3+3Q^2$.

Here we consider a linear dilaton background, 
so the string theory is weakly coupled in the region far from the 
singularity.
On the other hand, the string coupling constant
diverges near the singularity,
hence the perturbative approach is not reliable.
Thus we must add the Liouville potential to the worldsheet action of 
the Liouville theory in order to guarantee that
strings do not propagate into the region near the singularity.
But this additional term is actually the screening charge which
commutes with all the generators of $N=2$ superconformal algebra (3).
Thus although we cannot set the actual interaction to be vanish,
we can pursue all the manipulations like a free worldsheet CFT 
without the Liouville potential.
This situation is physically realized by taking the double scaling limit in
\cite{gk}.
This limit holds the mass of wrapped branes at finite value,
so we may not see the gauge symmetry enhancement, which is
characteristic phenomena at the singularity.

For the isolated ADE singularity, the Landau-Ginzburg theory with $W=F$
is nothing but the $N=2$ minimal models ($MM$) classified by ADE,
and have the central charge $c=\frac{3(N-2)}{N}$, where 
$N=k+2$ is the dual Coxeter number of the ADE groups. 
In this paper, we use both $N$ and $k$ in order to show the level of
$N=2$ minimal model, however it may be no confusion.

The condition for cancellation of conformal anomaly can be written as  
\begin{equation}
d\times \left(1+\frac{1}{2}\right)+
\frac{3N-6}{N}+3(1+Q^2)+11-26=0\label{crit},
\end{equation}
and then it is easy to determine the background charge $Q$ for each of
the cases $d=6,4,2$. 
In the case $d=6$, we obtain
\begin{equation}
Q=\sqrt{\frac{2}{N}}.
\end{equation}

For the case of singular K3 surface, 
the corresponding Landau-Ginzburg theories
are described by the following superpotential \cite{ov}
\begin{equation}
W_G=z^{-N}+F_G,
\end{equation}
where $G=ADE$ and $F_G$ is defined by (\ref{ade}).
These non-compact Landau-Ginzburg theories describe conformal
field theories with $c=6$, which can be reinterpreted by the coset
models 
\begin{equation}
\left(\frac{SL(2)_{N+2}}{U(1)}\times\frac{SU(2)_{N-2}}{U(1)}
\right)/{\bf Z}_N.
\end{equation}
The non-compact $z$-dependent piece, corresponding to the $SL(2)$ factor
in the coset plays a role to push up the central charge into the right value.
The equivalence between this non-compact Kazama-Suzuki model and the
$N=2$ Liouville theory was discussed in \cite{ov, es}.
It was pointed out that both theories are related by a kind of
$T$-duality \cite{ov}.

\subsection{Modular invariant partition function on a torus}
\hspace{5mm}
Let us consider the modular invariant partition function on a torus for
the above non-critical superstrings in the light-cone gauge
(${\bf R}^{d-2}\times{\bf R}_{\phi}\times S^1\times MM$).
The toroidal partition function factorizes into two parts
\begin{equation}
Z_0(\tau,\bar{\tau})\;Z_{GSO}(\tau,\bar{\tau}),
\end{equation}
where $Z_{GSO}$ contains the contributions 
on which GSO projection acts non-trivially,
and we denote the remaining part by $Z_0$.

The part $Z_0$ has only the contributions from the transverse
non-compact bosonic coordinates ${\bf R}^{d-2}\times {\bf R}_{\phi}$, or
the flat spacetime bosonic coordinates and the linear dilaton $\phi$.
The Liouville sector is a bit subtle because of the background charge.
We use the ansatz that only the normalizable states 
contribute to the partition function.
The normalizable spectrum in Liouville theory, 
in the sense of the delta function
normalization because the spectrum is continuous,
has the lower bound $h=Q^2/8$ \cite{kus2}.
This bound is nonzero, thus we must carefully handle the integration
over the zero-mode momentum.
However it turns out that
the resulting partition function of $\phi$ is effectively the 
same as that of a ordinary boson
because the effective value of the Liouville central charge $c_{{\rm
eff},L}$
is equal to 
\begin{equation}
c_{{\rm eff},L}\equiv (1+3Q^2)-24\times\frac{Q^2}{8}=1,
\end{equation}
which is independent of the background charge \cite{kus2}. 
Note that it is not clear whether
we should include the other modes.
However we do not concern with that point in this paper.

Thus we obtain $Z_0$ effectively as the contribution from $d-1$ free bosons  
\begin{equation}
Z_0(\tau,\bar{\tau})=\left(
\frac{1}{\sqrt{\tau_2}|\eta(\tau)|^2}\right)^{d-1}, \qquad
 \tau=\tau_1+i\tau_2,
\end{equation}
which is manifestly modular invariant.

The part $Z_{GSO}$ should be treated separately for $d=6,4,2$
due to the specific GSO projection.
We only mention the simplest case $d=6$, corresponding to singular K3 surface.

In order to specify the GSO projection, we have to consider the Fock
space of the bosonic circular space-time coordinates $Y$ 
constructed on the Fock vacuum $|p\rangle$, 
$\oint i \partial Y|p\rangle=p|p\rangle$.
The values of the momenta $p$ are chosen in
consistent with the GSO projection.
The conditions for the GSO projection on the $U(1)$ charge, 
which ensures the mutual locality with the space-time SUSY charges, 
are given by the following manner \cite{gkp}
\begin{eqnarray}
F+F_{MM}+\frac{m}{N}-pQ &\in& 2{\bf Z}+1, \qquad {\rm NS\: sector},\nonumber\\
F+F_{MM}+\frac{m}{N}-pQ &\in& 2{\bf Z}, \qquad\qquad {\rm R\:\; sector},
\label{gso}
\end{eqnarray}
where $F$ denotes the fermion number of 
${\bf R}^{d-2}\times({\bf R}_{\phi}\times S^1)$ sector and $F_{MM}$ 
denotes the fermion number of the minimal model
(For the notation of minimal model, see the Appendix A).

We compute the trace over the left-moving Hilbert space. 
For example, consider the NS sector with $F+F_{MM}\in 2{\bf Z}+1$.
The sum over the momenta becomes
\begin{equation}
\sum q^{\frac{1}{2}p^2}=\sum_n q^{\frac{N}{4}\left(2n+\frac{m}{N}\right)^2}
=\sum_n q^{N\left(n+\frac{m}{2N}\right)^2}=\theta_{m,N}(\tau).
\end{equation}
Then with the factors coming from oscillator modes and the minimal
modes, we obtain
\begin{equation}
\frac{1}{2}\left[
\left(\frac{\theta_3}{\eta}\right)^3 {\rm Ch}_{\ell, m}^{NS,(N-2)}
+\left(\frac{\theta_4}{\eta}\right)^3\widetilde{\rm Ch}_{\ell, m}^{NS,(N-2)}
\right]
\frac{\theta_{m,N}}{\eta},
\end{equation}
where $\eta, \theta_i \; (i=2,3,4)$ are the usual Dedekind, Jacobi
theta functions.
In this contribution, $\left(\frac{\theta_i}{\eta}\right)^2$ comes
from the ${\widehat {SO(4)}}_1$ character which is the contribution of
the fermionic fields in ${\bf R}^4$.
Additional contribution $\frac{\theta_i}{\eta}$ comes from the
contribution of fermionic fields in $N=2$ Liouville theory.
Almost in the same way, we can write down the whole contribution by 
\begin{eqnarray}
&&\frac{1}{2}\sum_{\ell=0}^{N-2}\sum_{m\in{\bf Z}_{2N}}
\Biggl[\theta_3^3\:{\rm Ch}_{\ell, m}^{NS,(N-2)}\left(
\theta_{m,N}+\theta_{m+N,N}\right)-
\theta_4^3\:\widetilde{\rm Ch}_{\ell, m}^{NS,(N-2)}
\left(\theta_{m,N}-\theta_{m+N,N}\right)\nonumber\\
&&\qquad\qquad\qquad
-\theta_2^3\:{\rm Ch}_{\ell, m}^{R,(N-2)}
\left(\theta_{m,N}+\theta_{m+N,N}\right)
\Biggr],
\end{eqnarray}
where $\widetilde{R}$ sector vanishes, and we have omitted the factor of 
$\eta$ function for simplicity.
Note that this sum counts each state twice due to the 
field identification for the character of minimal model.
In order to avoid this double counting, it is convenient to define
\begin{eqnarray}
F_{\ell}(\tau)\equiv\frac{1}{2}\sum_{m\in{\bf Z}_{2N}}
\theta_{m,N}\left(\theta_3^3\:{\rm Ch}_{\ell, m}^{NS,(N-2)}
-\theta_4^3\:\widetilde{\rm Ch}_{\ell, m}^{NS,(N-2)}
-\theta_2^3\:{\rm Ch}_{\ell, m}^{R,(N-2)}
\right)(\tau),\label{fl}
\end{eqnarray}
and construct modular invariants using this $F_{\ell}$.

Although we can read off the modular property of $F_{\ell}$
directly from the above definition,
it is convenient to introduce $F_{\ell}$ with $z$ dependence 
\begin{eqnarray}\label{z-dep}
F_{\ell}(\tau,z)&=&
\frac{1}{2}\sum_{m\in{\bf Z}_{2N}}
\theta_{m,N}(\tau,-2z/N)\\
&&\qquad\qquad\times \left(\theta_3^3\:{\rm Ch}_{\ell, m}^{NS,(N-2)}
-\theta_4^3\:\widetilde{\rm Ch}_{\ell, m}^{NS,(N-2)}
-\theta_2^3\:{\rm Ch}_{\ell, m}^{R,(N-2)}
-i\theta_1^3\:\widetilde{{\rm Ch}}_{\ell, m}^{R,(N-2)}
\right)(\tau,z).\nonumber 
\end{eqnarray}
Then due to the branching relation (\ref{br}), we can express $F_{\ell}$ in the
following form \cite{ov} 
\begin{equation}
F_{\ell}(\tau,z)=\frac{1}{2}\left(\theta_3^4-\theta_4^4-\theta_2^4+
\theta_1^4\right)
(\tau,z)
\;\chi_{\ell}^{(N-2)}(\tau,0),\label{t-dual}
\end{equation}
where $\chi_{\ell}^{(k)}$ denotes ${\widehat {SU(2)}}_k$ character of the
spin $\ell/2$ representation (\ref{su2}).
Thanks to this relation, we can easily find that
$F_{\ell}$ shows the same modular transformation property as the 
affine $SU(2)$ character.
Now we can construct the modular invariant partition function on a torus 
\begin{equation}
Z_{GSO}(\tau,\bar{\tau})=
\frac{1}{|\eta(\tau)|^8}
\sum_{\ell,\bar{\ell}=0}^{N-2}
N_{\ell,\bar{\ell}}\: F_{\ell}(\tau)\: F_{\bar{\ell}}(\bar{\tau}),\label{mod}
\end{equation}
where $N_{\ell,\bar{\ell}}$ is the Cappelli-Itzykson-Zuber matrix, which 
can be classified by ADE \cite{ciz}.
In this way, we can obtain the modular invariant classified by the ADE
groups corresponding to the singularity type of $X_n$ \cite{ov, ds}.
In this partition function, it appears 
a mass gap and the continuous spectrum above the gap due to the
Liouville theory.

Note that $F_{\ell}$ identically vanishes by virtue of the Jacobi's abstruse
identity in (\ref{t-dual}).
This is consistent with the existence of space-time supersymmetry.
Furthermore, the appearance of the ${\widehat {SU(2)}}$ character in
(\ref{t-dual}), and the standard ADE
classification of modular invariant corresponding to the type of
degeneration of K3 surface, which coincides exactly with the well-known
modular invariants of $SU(2)$ WZW theory,
are quite satisfactory pictures.
This originates from the following argument.
In some sense, we can relate the background of singular K3 surface to
a collection of NS5-branes by means of $T$-duality \cite{ov}.
Moreover, it was argued that the world-sheet CFT of superstrings on NS5
brane background contains the $SU(2)$ WZW theory in the near horizon 
regime \cite{chs}.

In the case of singular three- or four-fold with an isolated ADE
singularity,
we can similarly construct modular invariant partition function on
a torus using the ADE classification of modular invariant \cite{es}.
However the worldsheet interpretation of the results is not so much
clear as the singular K3 surface.
Except for the conifold, we do not know
the dual description by intersecting NS5 branes \cite{bsv}.

\section{Gepner models}
\hspace{5mm}
In this section we review the construction of modular invariant
partition function of Gepner models by the spectral flow method 
\cite{eoty}.
A characteristic feature of this method is that the space-time 
supersymmetry is manifest and the partition function on torus 
can be constructed by the block diagonal way.
As we will see in the next section, 
this block diagonal partition function 
is essential in order to see the proposed identifications.

We first mention the original work of Gepner \cite{g} and then 
give a brief review of spectral flow method. We only consider 
the smooth Calabi-Yau compactifications in this sections.
However the basic idea does not change even though one consider
Calabi-Yau manifolds with isolated ADE singularities.

\subsection{Spectral flow method}
\hspace{5mm}
Let us discuss type II string theory compactified on the smooth Calabi-Yau
manifolds of complex dimension $n$ 
($n=1$ for the torus, 2 for the K3 surface 
and 3 for the Calabi-Yau threefold). The transverse space in the
light-cone gauge is described by  
free bosons and free fermions.
To describe the internal space by exactly solvable CFT, 
one consider a tensor product of $r$
$N=2$ minimal models of level $k_1,\ldots , k_r$. 
In fact, we need certain conditions to construct the supersymmetric
string compactifications.
The cancellation of the trace anomaly requires that the central charges
of minimal models must add up to $3n$.
We further impose the projection that total $U(1)$ charges (sum of
the charges from the transverse SCFT and from the internal SCFT)
in both the left moving and right moving sector should be odd integers. 
Then we require the sector arraignment, which means that the left-moving
states 
and the right-moving states must be taken from the NS sectors of each
sub-theory 
or from the R sectors of each sub-theory and do not mix both sectors.
Gepner \cite{g} constructed the consistent modular invariant
partition functions which is
compatible with these conditions. 
This is so-called the $\beta$-method.
However, the result is too complicated and block diagonalization of
partition function can not be seen manifestly, so this
procedure is not suitable for our goal. 
Thus we give another so-called spectral flow method,
which gives the same result as the $\beta$-method.
  
A well-known feature of the $N=2$ algebra is the isomorphism of the
algebra under the continuous shift of the moding of the 
generators, i.e. under the spectral flow,
\begin{eqnarray}\label{flow}
L_n&\to& L_n+\eta\: J_n+{1\over 6}\:c\:\eta^2\: \delta_{n,0},\n\\
J_n&\to& J_n+{1\over 3}\:c\:\eta\:\delta_{n,0},\\
G_r^{\pm}&\to&G_{r\pm\eta}^{\pm},\n
\end{eqnarray}  
where $L_n$, $J_n$, $G_r^{\pm}$ are Virasoro, $U(1)$ current and
supercharge generators, respectively and $\eta$ is an arbitrary 
real parameter. 
The space-time supersymmetry transformation corresponds to the shift $\eta \to
\eta+{1\over 2}$, which exchanges NS sector for R sector.
Thus if the total Hilbert space is 
invariant under the shift $\eta \to\eta+{1\over 2}$, 
the supersymmetry is manifest. 
Further under the shift $\eta\to \eta +1$, NS sector comes back to
NS sector, however in general the states in NS sector are mapped onto the
different states. Therefore we repeat to 
operate the spectral flow until we return
to the original states. 

Partition function of Gepner models on which GSO 
projection acts non-trivially 
is expressed in terms of the characters of $N=2$ minimal model and 
free fermion.
For a given representation of $N=2$ minimal model, 
we can define the characters in each sector (see Appendix A).
Under the spectral flow with parameter $\eta={1 \o 2}$ 
or equivalently $z\to z+{\tau \o 2}$ with a factor 
$q^{c\o 24}\: y^{{c\over 6}}$, which comes from the shift of
zero mode in (\ref{flow}), 
the character in the NS sector becomes
\begin{equation}\label{eta=1/2}
q^{{c\o 24}}\:y^{{c\over 6}}\C^{NS,(k)}_{\ell,m}\left(\tau,z+{\tau \o 2}\right)
=\C^{R,(k)}_{\ell,m-1}(\tau, z)
\end{equation}  
and under the full shift $\eta =1$ or $z\to z+\tau$ with a factor 
$q^{c\o 6}\: y^{{c\over 3}}$, 
the character in the NS sector becomes
\begin{equation}\label{eta=1}
q^{{c\o 6}}\: y^{{c\over 3}}\C^{NS,(k)}_{\ell,m}\left(\tau,z+{\tau}\right)
=\C^{NS,(k)}_{\ell,m-2}(\tau, z),
\end{equation}
where we have the same expression in R sector.

Let us consider how to construct the partition function of Gepner
models by the spectral flow method.
We first define ``supersymmetric characters'' which is the building block of the partition function. 
We multiply all the characters in NS sector
which include the ground state $h=q=0$ , i.e. 
$\C_{0,0}^{NS,(k_1)}\ldots\C_{0,0}^{NS,(k_r)}$. Then we apply the $\eta=1$ 
spectral flow operations (\ref{eta=1}) until we obtain the original state. 
We denote these spectral flow invariant combination by $NS_0$. 
The graviton corresponds to $h=0$ state and $NS_0$ is called `graviton
orbit'. Under the modular
transformations $S:\tau\to -{1\over\tau}$,
we obtain a family of new spectral flow invariant orbits $NS_i$
(the range of $i$ depends on the models). 
We iterate this procedure until they transform among themselves under
the modular $S$ transformation:
\begin{equation}\label{S-trans}
NS_i\left(-{1\over \tau}\right)=\sum_i S_{ij}\: NS_j(\tau),
\end{equation}
where $S_{ij}$ is real $S$-matrix satisfying $S^2=1$.

Then the contribution from other sectors is obtained in a
straightforward way.
The orbits in the R sector can be obtained by the spectral flow
(\ref{eta=1/2}) 
and the orbits in the $\widetilde{\mbox{NS}}$ 
or the $\widetilde{\mbox{R}}$ sector, which are needed 
to close the orbits under the modular transformations 
$S:\tau\to-{1\o \tau}$ and $T:\tau\to\tau +1$, are obtained 
by the flow $z\to z+ {1\o 2}$.
The modular transformation matrix of these orbits is the same as that
of (\ref{S-trans}) \cite{eoty}.
Therefore we introduce the
supersymmetric character  
\begin{equation}
X_i\:(\tau, z)={1\over 2}\left\{
NS_i\left(\theta_3 \over \eta \right)^m
-{\widetilde {NS}}_i\left(\theta_4 \over \eta \right)^m
-R_i\left(\theta_2 \over \eta \right)^m
+{\widetilde R}_i\left(\theta_1 \over \eta \right)^m
\right\}(\tau, z),
\end{equation}
where $({\t\o \eta})^m$ come from the 
$\widehat{SO(2m)}_1$ characters with $m=4-n$ which is
the contribution of the spinor fields 
of the transverse flat space. 
This character is spectral flow invariant and therefore space-time
supersymmetry is manifest. 

Now we would like to construct the modular invariant partition function
on a torus.
Under the modular $T$ transformation, the supersymmetric
character is invariant up to a total phase factor. 
Under the modular $S$ transformation, $S$-matrix of the character is
identical to that of $NS_i$ (\ref{S-trans}).
Therefore we can easily construct the modular invariant partition
function.
We can define a particular diagonal matrix $D$ 
\begin{equation}
D_i={S_{0i}\over S_{i0}},
\end{equation}
satisfying
\begin{equation}
\sum_i \:S_{ij}\: D_i\: S_{ik}=D_j\:\delta_{jk}.
\end{equation}
Then the partition function on the torus is obtained by the following
bilinear modular invariant combination
\begin{equation}
Z=\sum_i D_i \: |X_i|^2.
\end{equation}
We can check that this gives the same partition function
constructed by the $\beta$-method. 
A characteristic feature of the spectral flow method is that if we want
to construct the supersymmetric characters and to know the modular
transformation of them, we have only to obtain the flow invariant orbits
of the NS sector and the modular invariance among themselves.

Finally we should comment on the reason for the block diagonalization of 
partition function. 
In general, if the theory has a certain 
enlarged algebra in the theory,
partition function is block diagonalized or fully
diagonalized in that algebra.
If we include the generators of spectral flow with 
$\eta =\pm 1$ to the original $N=2$ algebra,
we can extend it to enlarged algebra (in particular, in the case of K3 surface
the algebra becomes $N=4$ superconformal algebra).
Thus in this enlarged algebra, the partition function is block
diagonalized or fully diagonalized.

\section{$D_4,E_6,E_8$ modular invariants from tensor products}
\hspace{5mm}
Let us consider the $N=2$ Landau-Ginzburg theory in two dimensions \cite{lvw}.
Due to the singularity theory, the form of the superpotential
is classified by ADE, which correspond to $c<3$ unitary $N=2$ conformal
minimal models which have the same ADE classification as $SU(2)$ WZW
models \cite{ciz}.
The validity of this picture is checked by the equivalence of elliptic
genus \cite{w2, kyy}.
In particular, we concentrate on the $D_4,E_6, E_8$ modular invariants. 
These are very special modular invariants 
only with the block diagonal form, which may signal that the
representations in 
these partition functions form the reducible representation 
of original ${\widehat {SU(2)}}$ symmetry.
Originally we have a whole Hilbert space spanned by all the states with
$\ell=0,\dots ,k$ (spin $\ell/2$ representations), 
but the solutions tell us that 
there exist sensible physical system which have only the exponent
of $D_4,E_6, E_8$ groups.
In the case of $D_4$, the appearance of factor 2 is the specific feature.
Also, in $E_6, E_8$ case,
we can expect that combined with some larger symmetry, 
the partition function will be diagonalized \cite{bn}.
However this largerer symmetry hides the original relation
between $SU(2)$ WZW models and $N=2$ minimal
model by GKO coset construction \cite{gko}. 

In view of Landau-Ginzburg potential (\ref{ade}), 
we can expect that $D_4,E_6,E_8$ 
theories can be recaptured via $A_2\otimes A_2, A_2\otimes A_3,
A_2\otimes A_4$ Gepner models \cite{mvw}.
For $D_4$ case, we can rewrite the polynomial into the form $x^3+y^3$.
We should be able to see this correspondence at the level of the partition
functions.
However we can not make modular invariant partition function
using spectral flow invariants
only with $A_2\otimes A_2, A_2\otimes A_3, A_2\otimes A_4$ minimal
models.
If we try to construct the spectral flow orbit only with minimal models,
we encounter the bad $T$-transformation property 
due to the absence of integrality of $U(1)$ charge.
Of course, also in the case of smooth Calabi-Yau compactification classified
by ADE \cite{fkss}, we can expect above phenomena.
But due to the complexity, this problem has not been investigated.

Thus we wish to consider this identification in the singular Calabi-Yau
compactification.
In the present situation, we have originally have the negative power 
superpotential, which is somewhat difficult to tackle.
But we have replaced the negative term with the Liouville field,
so we may use the above standard logic. 
Moreover as you can see from previous sections, the spectral flow method is
quite more suitable with this situation than the $\beta$-method
in order to construct the tensor product theory. 
Furthermore, we can construct the explicit space-time
supersymmetric multiplet.
Here, we have the following two interests.
At first, we have an interest to reveal peculiar Hilbert space 
contained in block diagonal $D_4,E_6,E_8$ invariants.
In the second, this is the simplest setting to make the tensor product theory.
Our interest is the correspondence between these two objects.
Moreover this is the necessary consistency check if the partition function
reflects the singularity in spacetime.
We can make the similar construction 
for more complicated singularity in \cite{y},
but it is straightforward and may not be meaningful for the purpose in
this paper.

In the off-diagonal cases of $D_5, E_7$, 
the Landau-Ginzburg potentials are quartic.
Then we can rewrite the potential naively as in $D_4$ case.
However we cannot make the tensor product theory with the $c=9/4, 8/3$,
which corresponds to the value of central charge of $D_5, E_7$ theory.
For $D_{>6}$ case, we cannot rewrite the potential as in $D_4$ case.
Thus the other modular invariants $D_{n>4},E_7$ 
seem not to be constructed by the tensor product theory.

\subsection{$D_4$ case}
\hspace{5mm}
We wish to reproduce the $D_4$ modular invariant in (\ref{mod})
\begin{equation}  \label{d4}
Z_{GSO}^{D_4}=\frac{1}{|\eta|^8}
\left(
|F_0+F_4|^2+2|F_2|^2
\right),
\end{equation}
by the modular invariant of $A_2\otimes A_2$ Gepner model.

In the NS sector of $A_2$ minimal model at level 1, 
there are three irreducible representations.
We denote the characters associated with these
representations as 
\begin{equation}
A_1 = \C^{NS,(1)}_{0,0},\;\;
B_1 = \C^{NS,(1)}_{1,1},\;\;
C_1 = \C^{NS,(1)}_{1,-1}.
\end{equation}

Under the spectral flow with $\eta =1$, the above characters
change as follows 
\begin{equation}
A_1\to B_1\to C_1\to A_1\label{flow1},
\end{equation}
which is the diagrammatic expression of the flow (\ref{spectral}).

In order to construct the modular invariant partition function, we have
to specify
the condition of GSO projection.
The GSO projection is given by
\begin{eqnarray}
F+F_{MM_1}+F_{MM_2}+\frac{m_1+m_2}{3}-\frac{p}{\sqrt{3}} 
&\in& 2{\bf Z}+1, \qquad {\rm NS\: sector},\nonumber\\
F+F_{MM_1}+F_{MM_2}+\frac{m_1+m_2}{3}-\frac{p}{\sqrt{3}} 
&\in& 2{\bf Z}, \qquad\qquad {\rm R\:\; sector},
\end{eqnarray}
which is obvious generalization of (\ref{gso}) and $MM_1, MM_2$
represent two $A_2$ minimal models.
Then let us calculate the trace over the left-moving Hilbert space.
At first, consider the NS sector with $F+F_{MM_1}+F_{MM_2}\in 2{\bf
Z}+1$.
The sum over the momenta becomes 
\begin{equation}
\sum q^{\frac{1}{2}p^2}
=
\sum_n q^{\frac{3}{2}\left(2n+\frac{m_1+m_2}{3}\right)^2}
=
\sum_n q^{6\left(n+\frac{m_1+m_2}{6}\right)^2}
=
\theta_{2m_1+2m_2,6}\:(\tau).
\end{equation}
On the other hand, in the NS sector with $F+F_{MM_1}+F_{MM_2}\in 2{\bf
Z}$,
we obtain the following sum
\begin{equation}
\sum q^{\frac{1}{2}p^2}
=
\sum_n q^{\frac{3}{2}\left(2n+1+\frac{m_1+m_2}{3}\right)^2}
=
\sum_n q^{6\left(n+\frac{m_1+m_2+3}{6}\right)^2}
=
\theta_{2m_1+2m_2+6,6}\:(\tau).
\end{equation}

We have to make the spectral flow invariant orbit and the modular invariant
of the tensor product theory, then identify the pieces which coincide
with the block diagonal elements of the $D_4$ modular invariants (\ref{d4}).
In order to make the orbit in the manner as section 3,
we adopt the simple ansatz that the graviton orbit contains the term
\begin{equation}
A_1^2(\tau,z)\; \theta_{0,6}(\tau,-z/3),
\end{equation}
where we have inserted the $z$ dependence used in (\ref{z-dep}).
Then using the spectral flow (\ref{spectral}), we obtain the following
graviton orbit
\begin{equation}
NS_0=
A_1^2\;\left(\theta_{0,6}+\theta_{6,6}\right)
+
C_1^2\;\left(\theta_{2,6}+\theta_{8,6}\right)
+
B_1^2\;\left(\theta_{4,6}+\theta_{10,6}\right),
\end{equation}
where again we omit the factor of $\eta$ for simplicity.

Furthermore we can close the orbit of
this theory under $S$ modular transformation
using the additional spectral flow invariant orbit
\begin{equation}
NS_1
=
B_1C_1\; \left(\theta_{0,6}+\theta_{6,6}\right)
+
A_1B_1\; \left(\theta_{2,6}+\theta_{8,6}\right)
+
C_1A_1\; \left(\theta_{4,6}+\theta_{10,6}\right).
\end{equation}
Then the $S$ modular transformation is summarized by the 
following $S$ matrix
\begin{equation}
S_{ij}=\frac{1}{\sqrt{3}}
\left(
\begin{array}{cc}
1&2\\
1&-1
\end{array}
\right), \qquad i,j=0,1, \label{d4s}
\end{equation}
which acts on $NS_0, NS_1$ as in (\ref{S-trans}).

Now we can easily construct the modular invariant partition function 
as reviewed in section 3.
Let us define the supersymmetric characters
\begin{equation}
X_i(\tau,z)
=
\left(\theta_3^3 NS_i
-\theta_4^3 \widetilde{NS}_i
-\theta_2^3 R_i
-i\theta_1^3 \widetilde{R}_i
\right)(\tau,z), \qquad i=0,1,
\end{equation}
where  
$\widetilde{NS}_i, R_i, \widetilde{R}_i$
are obtained by the spectral flow (\ref{spectral}).
Using these supersymmetric characters, we can write down the modular
invariant partition function
\begin{equation}
Z_{GSO}^{A_2\otimes A_2}(\tau,\bar{\tau})=
\frac{1}{|\eta(\tau)|^8}
\left(|X_0|^2+2|X_1|^2
\right)(\tau,\bar{\tau}).
\end{equation}

How can we see the structure of $D_4$ ?
In fact, note that the $S$ matrix (\ref{d4s}) is equivalent to that for the 
$F_0+F_4, F_2$ pieces of $D_4$ theory in (\ref{d4}).
Thus we claim that the following equations  should hold
\begin{equation}
NS_0=F_0^{(NS)}+F_4^{(NS)},\qquad
NS_1=F_2^{(NS)},\label{equi}
\end{equation}
where $(NS)$ denotes the piece of NS sector in $F_{\ell}$ (\ref{fl}).
We have checked that the explicit $q$-expansion have the same form
in both sides.
Moreover one can check the equivalence in the other sector.
Thus we can say that we have reproduced the block diagonal elements of
$D_4$ modular invariant in terms of the spectral flow invariant orbits
of $A_2 \otimes A_2$ Gepner model, and observed the equivalence :
$Z_{GSO}^{D_4}=Z_{GSO}^{A_2 \otimes A_2}$.

In fact, the level 6 theta functions $\theta_{m,6}(\tau,0)$ are 
functionally independent for different $|m|$.
Thus we suspect that there must be the following
equivalence relation between the
coefficients of each theta function
\begin{eqnarray}\label{d4id}
{\rm Ch}_{0,0}^{NS,(4)}+{\rm Ch}_{4,0}^{NS,(4)}=A_1^2, 
&&{\rm Ch}_{2,0}^{NS,(4)}=B_1C_1,\nonumber\\
{\rm Ch}_{4,-4}^{NS,(4)}+{\rm Ch}_{4,2}^{NS,(4)}=C_1^2, 
&&{\rm Ch}_{2,2}^{NS,(4)}=A_1B_1,\\
{\rm Ch}_{4,-2}^{NS,(4)}+{\rm Ch}_{4,4}^{NS,(4)}=B_1^2, 
&&{\rm Ch}_{2,-2}^{NS,(4)}=C_1A_1.\nonumber 
\end{eqnarray}
We can give a complete proof for these identities between
the characters of minimal models at different levels. 
We show this in Appendix B.
The identity for other sectors can be obtained in a trivial way.

For the singular Calabi-Yau 3,4-folds,
the building block in \cite{es}
remains invariant under the spectral flow. 
These spectral flow invariant orbits are combined to make modular
invariant partition function using modular invariant of ${\widehat
{SU(2)}}$ and theta system. 
The additional modular invariant for the theta system 
is irrelevant to the relation between $D_4$ and $A_2 \otimes A_2$.
Thus the extension to the singular 3,4-fold
is straightforward due to the identity 
for minimal models (\ref{d4id}).

\subsection{$E_6$ case}
\hspace{5mm}
Let us consider the $E_6$ modular invariant in (\ref{mod})
\begin{equation}\label{e6}
Z_{GSO}^{E_6}=
\frac{1}{|\eta|^8}
\left(
|F_0+F_6|^2+|F_4+F_{10}|^2+|F_3+F_7|^2
\right).
\end{equation}
We wish to make the similar modular invariants
using the $A_2\otimes A_3$ Gepner model.

We label the characters of  
six irreducible representations in the NS sector for $A_3$ minimal model
at level two in the following way
\begin{equation}\label{char}
A_2 = {\rm Ch}^{NS,(2)}_{0,0},\;\;
B_2 = {\rm Ch}^{NS,(2)}_{2,2},\;\;
C_2 = {\rm Ch}^{NS,(2)}_{2,0},\;\;
D_2 = {\rm Ch}^{NS,(2)}_{2,-2},\;\;
E_2 = {\rm Ch}^{NS,(2)}_{1,1},\;\;
F_2 = {\rm Ch}^{NS,(2)}_{1,-1}.
\end{equation}
Now we use $A_2$ for both the label of level one minimal model and
that of the character in equation (\ref{char}), but there may be no confusion.
Under the spectral flow with $\eta =1$ in (\ref{spectral}), 
the above characters change as follows 
\begin{eqnarray}
&&A_2\to B_2\to C_2\to D_2\to A_2,\label{flow2}\\
&&\qquad\;\;\; E_2\to F_2\to E_2\label{flow3}.
\end{eqnarray}

Next we specify the condition of GSO projection in NS sector as follows
\begin{equation}
F+F_{MM_1}+F_{MM_2}
+\frac{m_1}{3}+\frac{m_2}{4}-\frac{p}{\sqrt{6}}\; \in\; 2{\bf Z}+1,
\end{equation}
where $MM_1, MM_2$ represent $A_2, A_3$ minimal models, respectively,
and $R$ sector has the obvious condition.
Then let us calculate the trace over the left-moving Hilbert space. 
Consider the NS sector with $F+F_{MM_1}+F_{MM_2}\in 2{\bf Z}+1$.
The sum over the momenta becomes
\begin{equation}
\sum q^{\frac{1}{2}p^2}=
\sum_n q^{\frac{6}{2}\left(2n+\frac{4m_1+3m_2}{12}\right)^2}
=\sum_n q^{12\left(n+\frac{4m_1+3m_2}{24}\right)^2}
=\theta_{4m_1+3m_2,12}\:(\tau).\label{mom1}
\end{equation}
On the other hand, in the NS sector with $F+F_{MM_1}+F_{MM_2}\in 2{\bf Z}$,
we obtain the following sum 
\begin{equation}
\sum q^{\frac{1}{2}p^2}=
\sum_n q^{\frac{6}{2}\left(2n+1+\frac{4m_1+3m_2}{12}\right)^2}
=\sum_n q^{12\left(n+\frac{4m_1+3m_2+12}{24}\right)^2}
=\theta_{4m_1+3m_2+12,12}\:(\tau).\label{mom2}
\end{equation}

We have to make the spectral flow invariant orbit and modular invariant
of the tensor product theory as $D_4$ case.
Again we adopt the simplest ansatz that the graviton orbit includes the term 
\begin{equation}
A_1\:A_2\:(\tau,z) \:\theta_{0,12}\:(\tau,-z/6),
\end{equation}
where the expected $z$-dependence has been included.
Then using the spectral flow (\ref{spectral}),
we obtain the following graviton orbit
\begin{eqnarray}
{\rm NS}_0&=&
A_1A_2\:\theta_{0,12}+
C_1D_2\:\theta_{2,12}+
B_1C_2\:\theta_{4,12}
+A_1B_2\:\theta_{6,12}\nonumber\\
&&+C_1A_2\:\theta_{8,12}+
B_1D_2\:\theta_{10,12}
+A_1C_2\:\theta_{12,12}
+C_1B_2\:\theta_{14,12}\nonumber\\
&&+B_1A_2\:\theta_{16,12}
+A_1D_2\:\theta_{18,12}
+C_1C_2\:\theta_{20,12}
+B_1B_2\:\theta_{22,12}.
\end{eqnarray}
Then we find that we can close the $S$ modular transformation using
additional spectral flow invariant orbits
\begin{eqnarray}
{\rm NS}_1& =& 
A_1C_2\:\theta_{0,12}+C_1B_2\:\theta_{2,12} 
+B_1A_2\:\theta_{4,12}+A_1D_2\:\theta_{6,12}\nonumber\\
&&+C_1C_2\:\theta_{8,12}+B_1B_2\:\theta_{10,12}
+A_1A_2\:\theta_{12,12}+C_1D_2\:\theta_{14,12}\nonumber\\
&&+B_1C_2\:\theta_{16,12}+A_1B_2\:\theta_{18,12}
+C_1A_2\:\theta_{20,12}+B_1D_2\:\theta_{22,12},\\
{\rm NS}_2& =& 
B_1F_2\:(\theta_{1,12}+\theta_{13,12})+A_1E_2\:(\theta_{3,12}+\theta_{15,12}) 
\nonumber\\
&&+C_1F_2\:
(\theta_{5,12}+\theta_{17,12})+B_1E_2\:(\theta_{7,12}+\theta_{19,12})
\nonumber\\
&&+A_1F_2\:
(\theta_{9,12}+\theta_{21,12})+C_1E_2\:(\theta_{11,12}+\theta_{23,12}).
\end{eqnarray}

Then $S$ transformation is summarized as the following $S$ matrices,
\begin{equation}
S_{ij}=\frac{1}{2}
\left(
\begin{array}{ccc}
1 & 1 & \sqrt{2}\\
1 & 1 & -\sqrt{2}\\
\sqrt{2} & -\sqrt{2} & 0
\end{array}
\right), \qquad\qquad i,j=0,1,2.\label{e6s}
\end{equation}

Then we can obtain the modular invariant partition function
using the supersymmetric characters
\begin{equation}
Z_{GSO}^{A_2\otimes A_3}(\tau,\bar{\tau})=\frac{1}{|\eta(\tau)|^8}
\left(|X_0|^2+|X_1|^2+|X_2|^2\right)(\tau,\bar{\tau}).
\end{equation}

Note that the $S$ matrix (\ref{e6s}) is equivalent to that for
the block diagonal pieces $F_0+F_6, F_4+F_{10}, F_3+F_7$
of $E_6$ modular invariant theory (\ref{e6}).
Thus we claim that the following equations should hold
\begin{equation}
NS_0=F_0^{(NS)}+F_6^{(NS)},
\quad NS_1=F_4^{(NS)}+F_{10}^{(NS)},
\quad NS_2=F_3^{(NS)}+F_7^{(NS)},
\end{equation}
where $(NS)$ denotes the contribution of NS sector in (\ref{fl}).
Again we have checked that the explicit $q$ expansion have the same form.
Also we can check the equivalence in the other sector
via the explicit $q$-expansion and modular property.
Thus we have reproduced the block diagonal elements of $E_6$ modular
invariants in terms of the spectral flow invariant orbits by
$A_1\otimes A_2$ tensor products.

Note that the pattern of multiplication of theta function in
each spectral flow invariant orbit is
different between $NS_0, NS_1$ and $NS_2$. 
The number of element is different
in two flow invariant from minimal models, using
(\ref{flow1}), (\ref{flow2}) or (\ref{flow1}), (\ref{flow3}). 
Also $F_0+F_6, F_4+F_{10}$ in (\ref{e6}) 
do not close by itself under field identification in
minimal model, but $F_3+F_7$ closes by itself.

In fact, the level-$12$ theta functions $\theta_{m,12}(\tau,0)$
are functionally independent for different $|m|$, thus
we can expect that there should be the equivalence relation between the 
characters in minimal models, such as
\begin{eqnarray}
{\rm Ch}_{0,0}^{NS,(10)}+{\rm Ch}_{6,0}^{NS,(10)}&=& A_1A_2.\label{e6id}
\end{eqnarray}
The other equations like this are easily obtained.
Furthermore we can prove the identity exactly (Appendix B). 

\subsection{$E_8$ case}
\hspace{5mm}
We can proceed in the same way as the $D_4, E_6$ case.
In this case, we wish to 
construct modular invariants of $A_2\otimes A_4$ tensor 
product, and reproduce the structure of $E_8$ modular invariant theory
in (\ref{mod})
\begin{equation}
Z_{GSO}^{E_8}
=
\frac{1}{|\eta|^8}
\left(
|F_0+F_{10}+F_{18}+F_{28}|^2+|F_6+F_{12}+F_{16}+F_{22}|^2 
\right)
\label{e8}.
\end{equation}

First we label the NS characters of ten irreducible representations 
of $A_4$ minimal model at level 3 as follows
\begin{eqnarray}
&&A_3= {\rm Ch}^{NS,(3)}_{0,0},\;\;
B_3 = {\rm Ch}^{NS,(3)}_{3,3},\;\;
C_3 = {\rm Ch}^{NS,(3)}_{3,1},\;\;
D_3 = {\rm Ch}^{NS,(3)}_{3,-1},\;\;
E_3 = {\rm Ch}^{NS,(3)}_{3,-3},\;\;\nonumber\\
&&
\:F_3 = {\rm Ch}^{NS,(3)}_{1,1},\;\;
G_3 = {\rm Ch}^{NS,(3)}_{1,-1},\;\;
H_3 = {\rm Ch}^{NS,(3)}_{2,2},\;\;
I_3 = {\rm Ch}^{NS,(3)}_{2,0},\;\;
J_3 = {\rm Ch}^{NS,(3)}_{2,-2}.\;\;
\end{eqnarray}
Then we can summarize the action of spectral flow in the following manner
\begin{eqnarray}
&&A_3 \to B_3 \to C_3 \to D_3 \to E_3 \to A_3,\label{flow4}\\
&&\:F_3 \to G_3 \to H_3 \to I_3 \to J_3 \to F_3\label{flow5}, 
\end{eqnarray}
and there exist two naive spectral flow invariant orbits 
using (\ref{flow1}), (\ref{flow4}), (\ref{flow5}) in $A_2\otimes A_4$ theory.

The GSO projection in the NS sector is given by
\begin{equation}
F+F_{MM_1}+F_{MM_2}
+\frac{m_1}{3}+\frac{m_3}{5}-\frac{p}{\sqrt{15}}\; \in\; 2{\bf Z}+1,
\end{equation}
where we denote $A_2, A_4$ minimal models by $MM_1, MM_2$ respectively,
and R sector has the similar condition. 
Again, consider the NS sector with $F+F_{MM_1}+F_{MM_2}\in 2{\bf Z}+1$.
The sum over the momenta becomes
\begin{equation}
\sum q^{\frac{1}{2}p^2}=
\sum_n q^{\frac{15}{2}\left(2n+\frac{5m_1+3m_3}{30}\right)^2}
=\sum_n q^{30\left(n+\frac{5m_1+3m_3}{60}\right)^2}
=\theta_{5m_1+3m_3,30}\:(\tau).
\end{equation}
For the NS setor with $F+F_{MM_1}+F_{MM_2}\in 2{\bf Z}$, we obtain the 
following sum
\begin{equation}
\sum q^{\frac{1}{2}p^2}=
\sum_n q^{\frac{30}{4}\left(2n+1+\frac{5m_1+3m_3}{30}\right)^2}
=\sum_n q^{30\left(n+\frac{5m_1+3m_3+30}{60}\right)^2}
=\theta_{5m_1+3m_3+30,30}\:(\tau).
\end{equation}

Then the same way as previous subsection, 
we can make the following graviton orbit
\begin{eqnarray}
{\rm NS}_0& =& 
A_1A_3\:(\theta_{0,30}+\theta_{30,30})+C_1E_3\:(\theta_{2,30}+\theta_{32,30})
+B_1D_3\:(\theta_{4,30}+\theta_{34,30})\nonumber\\
&&+A_1C_3\:(\theta_{6,30}+\theta_{36,30})
+C_1B_3\:(\theta_{8,30}+\theta_{38,30})+B_1A_3\:(\theta_{10,30}+\theta_{40,30})
\nonumber\\
&&+A_1E_3\:(\theta_{12,30}+\theta_{42,30})+C_1D_3\:
(\theta_{14,30}+\theta_{44,30})
+B_1C_3\:(\theta_{16,30}+\theta_{46,30})
\nonumber\\
&&+A_1B_3\:(\theta_{18,30}+\theta_{48,30})
+C_1A_3\:(\theta_{20,30}+\theta_{50,30})+B_1E_3\:
(\theta_{22,30}+\theta_{52,30})
\nonumber\\
&&+A_1D_3\:(\theta_{24,30}+\theta_{54,30})+C_1C_3\:
(\theta_{26,30}+\theta_{56,30})
+B_1B_3\:(\theta_{28,30}+\theta_{58,30}).
\end{eqnarray}
Then we can close the system under $S$ transformation using the
additional spectral flow orbit
\begin{eqnarray}
{\rm NS}_1& =& 
A_1I_3\:(\theta_{0,30}+\theta_{30,30})+C_1H_3\:(\theta_{2,30}+\theta_{32,30})
+B_1G_3\:(\theta_{4,30}+\theta_{34,30})\nonumber\\
&&+A_1F_3\:(\theta_{6,30}+\theta_{36,30})
+C_1J_3\:(\theta_{8,30}+\theta_{38,30})+B_1I_3\:(\theta_{10,30}+\theta_{40,30})
\nonumber\\
&&+A_1H_3\:(\theta_{12,30}+\theta_{42,30})+C_1G_3\:
(\theta_{14,30}+\theta_{44,30})
+B_1F_3\:(\theta_{16,30}+\theta_{46,30})
\nonumber\\
&&+A_1J_3\:(\theta_{18,30}+\theta_{48,30})
+C_1I_3\:(\theta_{20,30}+\theta_{50,30})+B_1H_3\:
(\theta_{22,30}+\theta_{52,30})
\nonumber\\
&&+A_1G_3\:(\theta_{24,30}+\theta_{54,30})+C_1F_3
\:(\theta_{26,30}+\theta_{56,30})
+B_1J_3\:(\theta_{28,30}+\theta_{58,30}).
\end{eqnarray}
Notice that $NS_0, NS_1$ uses the spectral flow invariants  
$(\ref{flow1}), (\ref{flow4})$ and $(\ref{flow1}), (\ref{flow5})$,
respectively. 
The $S$ matrix for these orbits
coincides with that for the block diagonal term 
$F_0+F_{10}+F_{18}+F_{28}, F_6+F_{12}+F_{18}+F_{22}$
in $E_8$ invariants (\ref{e8})
\begin{equation}
S_{ij}=\frac{2}{\sqrt{5}}
\left(
\begin{array}{cc}
\frac{\sqrt{10-2\sqrt{5}}}{4}&\frac{\sqrt{10+2\sqrt{5}}}{4}\\
\frac{\sqrt{10+2\sqrt{5}}}{4}&-\frac{\sqrt{10-2\sqrt{5}}}{4}
\end{array}
\right), \qquad i,j=0,1.
\end{equation}
Thus we can write down modular invariant partition function 
using supersymmetric characters
\begin{equation}
Z_{GSO}^{A_2\otimes A_4}(\tau,\bar{\tau})=\frac{1}{|\eta(\tau)|^8}
\left(|X_0|^2+|X_1|^2
\right)(\tau,\bar{\tau}).
\end{equation}

In the same way as previous cases,
we can write down the following relation.
\begin{eqnarray}
&&NS_0=F_0^{(NS)}+F_{10}^{(NS)}+F_{18}^{(NS)}+F_{28}^{(NS)},\label{e81}\\
&&NS_1=F_6^{(NS)}+F_{12}^{(NS)}+F_{16}^{(NS)}+F_{22}^{(NS)},\label{e82}
\end{eqnarray}
where $(NS)$ denotes the contribution from NS sector in (\ref{fl}).
We have compared explicit $q$ expansions in both side and checked
the equivalence.
Thus in the same sense as the previous subsections, we have succeeded 
to rewrite the block diagonal elements in $E_8$ invariants
in terms of the spectral flow invariant orbits of $A_2\otimes A_4$ theory.
Compared with $E_6$ case where two naive spectral flow 
orbits in minimal model
divided into three orbits in whole theory, we have only two
orbits in whole system related to each orbits in minimal models. 
This is the same manner as in $D_4$ case. 

Moreover using the fact that
the level-$30$ theta functions $\theta_{m,30}(\tau,0)$
are functionally independent for different $|m|$, 
we can read off the identities among the characters in minimal models
contained in (\ref{e81}), (\ref{e82}), such as
\begin{equation}
{\rm Ch}_{0,0}^{NS,(28)}+{\rm Ch}_{10,0}^{NS,(28)}+{\rm Ch}_{18,0}^{NS,(28)}
+{\rm Ch}_{28,0}^{NS,(28)}
=
A_1A_3.\label{e8id}
\end{equation}
All the relation like this are easily obtained. 
Again we can give a exact proof of the identity, see Appendix B.

\section{Conclusion and discussion}
\hspace{5mm}
In this paper, we have studied the toroidal partition functions 
of non-critical superstring theory on ${\bf R}^{d-1,1}\times ({\bf
R}_{\phi}\times {\bf S}^1) \times M_{D_4,E_6,E_8}$, which is conjectured to
give the dual description of Calabi-Yau manifolds with 
the ADE singularity in the 
decoupling limit.
The ADE classification of modular invariants associated to the type
of Calabi-Yau singularities suggests that the natural reinterpretation
of $D_4,E_6, E_8$ theory via Gepner models of $A_2 \otimes A_2,
A_2 \otimes A_3, A_2 \otimes A_4$.
Strategy of the spectral flow invariant 
orbit has given more natural framework 
to the singular Calabi-Yau compactification 
than the conventional smooth Calabi-Yau compactification.
Moreover we have obtained the identities among the characters in the
minimal models at different levels.
Maybe the existence of these identities were implicitly known in the work
\cite{mvw}.
But in the present more realistic situation than only the minimal
models, we have been able to obtain the relations more naturally.
Furthermore we have given the complete proof of the identities.
Our work gives the basic consistency checks on the use of
Landau-Ginzburg theory for the singular Calabi-Yau compactification. 

The characters of minimal models are defined by taking care of
all the null states.
Thus our identity among the characters of minimal models
at different levels may seem to be
rather non-trivial.
However, in CFTs, 
we often encounter the phenomena that we can obtain the non-trivial
relation in some model by imposing the larger symmetry.
There would be some interest to 
investigate the precise structure of the identity between
each representations along the null field construction \cite{km}.

Here we pose the unresolved problem.
We can calculate the elliptic genus of the singular Calabi-Yau manifold
using the CFTs.
It turns out that the elliptic genus vanishes \cite{y}.
This fact is reflected in the following observation.
For example, in the case of K3 surface with the isolated ADE singularity,
we cannot reproduce any nontrivial Hodge number
\cite{abfgz} of the corresponding ALE space.
Thus the CFT system really does not respect the geometry of the 
ALE spaces.
Only the exception is the case of conifold \cite{mi} where
extra Hodge number has been appeared, it was claimed that
it correspond to an additional massless soliton as in \cite{s}.
On the other hand in the smooth CFTs, 
D-brane wrapped around a collapsing cycle becomes a
fractional brane \cite{ddg} with a finite mass.
Thus perturbative description is reliable at least if the string
coupling is small and the mass of fractional brane is large.
Then it would be unreasonable to claim that in the singular CFTs
the partition function includes the extra massless mode.
A sensible interpretation on the whole phenomena is not clear.

Moreover it would be interesting to 
consider the boundary states in these backgrounds
like \cite{ooy, rs}, and investigate the relation to Seiberg-Witten
theory as in \cite{l}.
Then, only the nontrivial part would be the
construction and interpretation
of boundary states for the Liouville sector \cite{li}.

\section*{Acknowledgments} 
\hspace{5mm}
M. Naka is grateful to T. Eguchi, T. Kawai, Y. Matsuo,
S. Mizoguchi, Y. Sugawara and S.-K. Yang for useful discussions. 
We would like to thank T. Takayanagi, T. Uesugi and S. Yamaguchi 
for helpful correspondence. 
Also M. Naka in particular thank S. Mizoguchi and his theory group 
for the kind hospitality at KEK while the part of this work was carried out.
The research of M. Naka is supported by JSPS
Research Fellowships for Young Scientists.
\newpage
\section*{Appendix A Convention of Conformal Field Theory}
\hspace{5mm}
In this appendix, we summarize the notation and collect the formulas used
in this paper.
We set $q=e^{2\pi i \tau}$ and $y=e^{2\pi i z}$.

\subsection*{1. Theta functions}
\hspace{5mm}
Jacobi theta functions are defined by
\begin{eqnarray}
&&\theta_1(\tau,z)=i\sum_{n=-\infty}^{\infty}
(-1)^nq^{\frac{1}{2}\left(n-\frac{1}{2}\right)^2}y^{n-\frac{1}{2}}
=2\:q^{1\o 8}\:
\sin{(\pi z)}
\prod_{m=1}^{\infty}(1-q^m)(1-yq^m)(1-y^{-1}q^m),\nonumber\\
&&\theta_2(\tau,z)=\sum_{n=-\infty}^{\infty}
q^{\frac{1}{2}\left(n-\frac{1}{2}\right)^2}y^{n-\frac{1}{2}}
=2\:q^{1\o 8}\:
\cos{(\pi z)}
\prod_{m=1}^{\infty}(1-q^m)(1+yq^m)(1+y^{-1}q^m),\nonumber\\
&&\theta_3(\tau,z)=\sum_{n=-\infty}^{\infty}q^{\frac{n^2}{2}}y^n
=\prod_{m=1}^{\infty}(1-q^m)(1+yq^{m-\frac{1}{2}})(1+y^{-1}q^{m-\frac{1}{2}}),
\nonumber\\
&&\theta_4(\tau,z)=\sum_{n=-\infty}^{\infty}(-1)^n
q^{\frac{n^2}{2}}y^n
=\prod_{m=1}^{\infty}(1-q^m)(1-yq^{m-\frac{1}{2}})(1-y^{-1}q^{m-\frac{1}{2}}).
\end{eqnarray}
For a positive
integer $k$, theta function of level $k$ is defined by
\begin{equation}
\theta_{m,k}(\tau,z)=\sum_{n=-\infty}^{\infty}
q^{k\left(n+\frac{m}{2k}\right)^2}
y^{k\left(n+\frac{m}{2k}\right)},
\end{equation}
where $m\in {\bf Z}_{2k}$.
We can rewrite the Jacobi theta functions in terms of the theta
function of level 2
\begin{eqnarray}
i\theta_1=\theta_{1,2}-\theta_{3,2},\qquad &&
\theta_2=\theta_{1,2}+\theta_{3,2},\n\\
\theta_3=\theta_{0,2}+\theta_{2,2},\qquad &&
\theta_4=\theta_{0,2}-\theta_{2,2}.
\end{eqnarray}
Dedekind $\eta$ function is represented as 
\begin{equation}
\eta(\tau)=q^{\frac{1}{24}}\prod_{n=1}^{\infty}(1-q^n).
\end{equation}

\subsection*{2. Characters of $N=2$ minimal model}
\hspace{5mm}
There is the discrete series of unitary representations of $N = 2$
superconformal algebra with
$c <3$, in fact with $c={3k \o k+2}$ ($k=N-2=1,2,3,\ldots$).
Based on these representations, one can construct families 
of conformal field theories known as $N = 2$ minimal models.
Their highest weight states are characterized by
conformal weight $h$ and the $U(1)$ charge $q$:
\begin{equation}
h^{\ell,s}_{m}={\ell(\ell+2)-m^2 \over 4(k+2)}+{s^2 \over 8},\qquad
q^{\ell,s}_{m}={m\over k+2}-{s\over 2}~,
\end{equation}
where $\ell\in\{0,\ldots,k\}$, $|m-s|\leq \ell$, $s\in\{-1,0,1,2\}$
and $\ell+m+s\equiv0$ mod 2. This range of $(\ell,m,s)$ is called `standard range'.

The discrete representations of $N=2$ algebra are 
related to the ${\widehat {SU(2)}}_k$ representations.
The character of ${\widehat {SU(2)}}_k$ with the spin ${\ell\o 2}$ 
($0\leq \ell\leq k$) representation is defined by
\begin{equation}
\chi_{\ell}^{(k)}(\tau,z)=\frac{\theta_{\ell+1,k+2}-\theta_{-\ell-1,k+2}}
{\theta_{1,2}-\theta_{-1,2}}(\tau,z)
:=\sum_{m\in{\bf Z}_{2k}}c_m^{\ell}(\tau)\:\theta_{m,k}
(\tau,z),\label{su2}
\end{equation}
where we refer to the coefficient $c_m^{\ell}(\tau)$ as string
function. String function has the following properties :
$c_m^{\ell}=c_{-m}^{\ell}=c_{m+2k}^{\ell}=c_{m+k}^{k-\ell}\;\;{\rm and}\;\;
c_m^{\ell}=0 \;\;{\rm unless}\;\; \ell+m\equiv 0\;\; ({\rm mod} \; 2)$.

On the other side, 
the character of $N=2$ representation 
labeled by $(\ell, m, s)$ is defined by 
$\chi^{\ell}_{m,s}(\tau, z)=\T_{{\cal H}^{\ell}_{m,s}}\:q^{L_0-{c\over 24}}\:
y^{J_0}$.
The explicit formula of $N=2$ character is obtained through the
branching relation \cite{g,kyy}
\begin{equation}
\chi_{\ell}^{(k)}(\tau,w)\:\t_{s,2}(\tau,w-z)=\sum_{m=-k-1}^{k+2}
\chi^{\ell,s}_m(\tau,z)\:\t_{m,k+2}\left(\tau,w-{2z\o k+2}\right)\,,\label{br}
\end{equation}
and is given by \cite{ry}
\begin{equation}
\chi_{m}^{\ell,s}(\tau,z)=
\sum_{r \in {\bf Z}_k} c_{m-s+4r}^{\ell}(\tau)\:\theta_{2m+(k+2)(-s+4r),2k(k+2)}
\left(\tau,\frac{z}{k+2}\right) .
\end{equation}
This character is actually defined in the `extended range' 
\begin{equation}\label{ext}
\ell\in\{0,\ldots,k\},\quad 
m\in {\bf Z}_{2k+4},\quad s\in {\bf Z}_4 \quad{\rm and} \quad
\ell +m+s\equiv 0 \;\; {\rm mod} \;\;2. 
\end{equation}
However, since the character has the following properties
\begin{equation}
\chi_m^{\ell,s}=\chi_{m+2k+4}^{\ell,s}=\chi_m^{\ell,s+4}
=\chi_{m+k+2}^{k-\ell,s+2}\;\;{\rm and}\;\;
\chi_m^{\ell,s}=0 \;\; {\rm unless} 
\;\; \ell+m+s\equiv 0 \;\; {\rm mod}\; 2~,
\end{equation}
we can always bring the range of $(\ell,m,s)$ into the standard range.

The characters of $N=2$ minimal model of level $k$ are 
defined by
\begin{eqnarray}
&&\C_{\ell, m}^{NS,(k)}(\tau ,z)
=\chi_m^{\ell,0}(\tau ,z)+\chi_m^{\ell,2}(\tau ,z),\quad
\widetilde{\C}_{\ell, m}^{NS,(k)}(\tau ,z)
=\chi_m^{\ell,0}(\tau ,z)-\chi_m^{\ell,2}(\tau ,z),\n\\
&&\C_{\ell, m}^{R,(k)}(\tau ,z)
=\chi_m^{\ell,1}(\tau ,z)+\chi_m^{\ell,3}(\tau ,z),\quad\;\;
\widetilde{\C}_{\ell, m}^{R,(k)}(\tau ,z)
=\chi_m^{\ell,1}(\tau ,z)-\chi_m^{\ell,3}(\tau ,z).
\end{eqnarray}
The explicit formula of the character 
in NS sector is represented as an
infinite product form \cite{d}
\begin{equation}
{\C}_{\ell, m}^{NS,(k)}(\tau,z)=
q^{h^{(NS)}_{\ell, m}-{c\o 24}}
\: y^{q^{(NS)}_{\ell,m}}
\:\prod_{n=1}^{\infty}
{(1+yq^{n-1/2})(1+y^{-1}q^{n-1/2})\o (1-q^n)^2}\Gamma_{\ell,m}^{(N)},\label{inf}
\end{equation}
where $N=k+2$ and
$ h^{(NS)}_{\ell, m}=h^{\ell,0}_m$, $ q^{(NS)}_{\ell,m}=q^{\ell, 0}_m$,
\begin{displaymath}
\Gamma_{\ell,m}^{(N)}=\prod_{n=1}^{\infty}
\frac{(1-q^{Nn+\ell+1-N})(1-q^{Nn-\ell-1})(1-q^{Nn})^2}
{(1+yq^{Nn-\frac{\ell+m+1}{2}})
(1+y^{-1}q^{Nn+\frac{\ell+m+1}{2}-N})
(1+y^{-1}q^{Nn-\frac{\ell-m+1}{2}})
(1+yq^{Nn+\frac{\ell-m+1}{2}-N})}.
\end{displaymath}

\subsection*{3. Modular transformations}
\hspace{5mm}
For simplicity, we use the following abbreviations : 
$\theta_{m,k}(\tau)\equiv\theta_{m,k}(\tau,0)$, 
$\chi_{m}^{\ell,s}(\tau)\equiv\chi_{m}^{\ell,s}(\tau,0)$.
Under the modular transformation $S:\tau \to -1/\tau$, the
characters defined above transform as 
\begin{eqnarray}
\chi_{\ell}^{(k)}(-1/\tau)&=&
\sum_{\ell'=0}^kS_{\ell\ell'}^{(k)}\:\chi_{\ell'}^{(k)}(\tau),\\
\theta_{m,k}(-1/\tau)&=&
\sqrt{-i\tau}
\sum_{m'\in{\bf Z}_{2k}}
\widetilde{S}_{mm'}^{(k)}\:\theta_{m',k}(\tau),\\
\chi_{m}^{\ell,s}(-1/\tau)&=&
\sum_{\ell',m',s'}S_{\ell\ell'}^{(k)}\:
\widetilde{S}_{mm'}^{(k+2)\dagger}\:
\widetilde{S}_{ss'}^{(2)}
\chi_{m'}^{\ell's'}(\tau),
\end{eqnarray}
where $\sum_{\ell',m',s'}$ denotes the summation over 
the extended range (\ref{ext}).
The modular transformation matrices of the characters are given by
\begin{eqnarray}
S_{\ell\ell'}^{(k)}&=&\sqrt{\frac{2}{k+2}}\:
\sin{\pi\frac{(\ell+1)(\ell'+1)}{k+2}},\\
\widetilde{S}_{mm'}^{(k)}&=&\frac{1}{\sqrt{2k}}\:e^{-2\pi i\frac{mm'}{2k}}.
\end{eqnarray}
Under the modular transformation $T:\tau\to\tau +1$, the characters
transform as 
\begin{eqnarray}
\chi_{\ell}^{(k)}(\tau+1)&=&e^{2\pi i\left[
\frac{\ell(\ell+2)}{4(k+2)}-\frac{c}{24}\right]}\:
\chi_{\ell}^{(k)}(\tau),\\
\theta_{m,k}(\tau +1)&=&e^{2\pi i\frac{m^2}{4k}}\:\theta_{m,k}(\tau),\\
\chi_{m}^{\ell,s}(\tau+1)&=&
e^{2\pi i\left[h^{\ell,s}_m-{c\o 24}\right]}\:\chi_{m}^{\ell,s}(\tau),
\end{eqnarray}
where $c={3k\o k+2}$. 

If we want to know how the characters transform under the spectral
flow, we have only to know the properties of the characters 
under the shift of parameter $z$. 
Then the characters $\chi^{\ell,s}_{m}(\tau, z)$ 
and $\theta_{m,N}(\tau,-2z/N)$
transform as
\begin{eqnarray}\label{spectral}
\chi^{\ell,s}_{m}\left(\tau, z+\frac{\tau}{2}\right)
&=& q^{-{c\o 24}}\:y^{-{c\over 6}}\:
\chi^{\ell,s-1}_{m-1}(\tau, z),\nonumber\\
\chi^{\ell,s}_{m}(\tau, z+\tau)&=& q^{-{c\o 6}}\:y^{-{c\over 3}}\:
\chi^{\ell,s}_{m-2}(\tau, z),\nonumber\\
\chi^{\ell,s}_{m}\left(\tau, z+\frac{1}{2}\right)
&=& (-i)^s \:e^{{i\pi m \over N}}\:
\chi^{\ell,s}_{m}(\tau, z),\nonumber\\
\theta_{m,N}\left(\tau,-{2(z+\tau) \o N}\right)&=&q^{-{1\o N}}\:y^{-{2\o N}}\:
\theta_{m-2,N}\left(\tau,-{2z\o N}\right),\\
\theta_{m,N}\left(\tau,-{2(z+\tau/2)\o N}\right)
&=&q^{-{1\o 4N}}\:y^{-{1\o N}}\:
\theta_{m-1,N}\left(\tau,-{2z\o N}\right),\nonumber\\
\theta_{m,N}\left(\tau,-{2(z+1/2)\o N}\right)&=&e^{-{i\pi m\o N}}\:
\theta_{m-1,N}\left(\tau,-{2z\o N}\right),\nonumber
\end{eqnarray}
where $N=k+2$.
\section*{Appendix B $\;$ Proof of identities between minimal models}
\hspace{5mm}
In this appendix, we give exact proofs of the identity 
(\ref{d4id}), (\ref{e6id}), (\ref{e8id}) among characters of minimal models.
The other identity can be proven along the same line.

\subsection*{1. $D_4$ case}
\hspace{5mm}
Let us consider the first identity in (\ref{d4id}).
We use the infinite product representation of minimal characters at
level $4,1$ (\ref{inf}) 
\begin{eqnarray}
{\rm Ch}_{0,0}^{NS,(4)}(\tau)&=&q^{-\frac{1}{12}}\:
\prod_{n=1}^{\infty}
\frac{\left(1+q^{n-\frac{1}{2}}\right)^2}{\left(1-q^n\right)^2}\cdot
\frac{\left(1-q^{6n-5}\right)\left(1-q^{6n-1}\right)\left(1-q^{6n}\right)^2}
{\left(1+q^{6n-\frac{1}{2}}\right)^2\left(1+q^{6n-\frac{11}{2}}\right)^2},
\nonumber\\
{\rm Ch}_{4,0}^{NS,(4)}(\tau)&=&q^{\frac{11}{12}}\:
\prod_{n=1}^{\infty}
\frac{\left(1+q^{n-\frac{1}{2}}\right)^2}{\left(1-q^n\right)^2}\cdot
\frac{\left(1-q^{6n-5}\right)\left(1-q^{6n-1}\right)\left(1-q^{6n}\right)^2}
{\left(1+q^{6n-\frac{5}{2}}\right)^2\left(1+q^{6n-\frac{7}{2}}\right)^2},
\nonumber\\
{\rm Ch}_{0,0}^{NS,(1)}(\tau)&=&q^{-\frac{1}{24}}\:
\prod_{n=1}^{\infty}
\frac{\left(1+q^{n-\frac{1}{2}}\right)^2}{\left(1-q^n\right)^2}\cdot
\frac{\left(1-q^{3n-2}\right)\left(1-q^{3n-1}\right)\left(1-q^{3n}\right)^2}
{\left(1+q^{3n-\frac{1}{2}}\right)^2\left(1+q^{3n-\frac{5}{2}}\right)^2},
\end{eqnarray}
where we set $z=0$ for simplicity.
Dividing by $\prod_{n=1}^{\infty}\left(1+q^{n-\frac{1}{2}}\right)^2/
\left(1-q^n\right)^2$, we can write down the identity (\ref{d4id}) : 
${\rm Ch}_{0,0}^{NS,(4)} \: {\rm Ch}_{4,0}^{NS,(4)} 
={\rm Ch}_{0,0}^{NS,(1)}$ as follows
\begin{eqnarray}\label{proc}
&&\prod_{n=1}^{\infty}\left(1-q^{6n-5}\right)
\left(1-q^{6n-1}\right)\left(1-q^{6n}\right)^2\nonumber\\
&&\times
\left[
\prod_{n=1}^{\infty}
\frac{1}
{\left(1+q^{6n-\frac{1}{2}}\right)^2\left(1+q^{6n-\frac{11}{2}}\right)^2}
+q\:\prod_{n=1}^{\infty}
\frac{1}
{\left(1+q^{6n-\frac{5}{2}}\right)^2\left(1+q^{6n-\frac{7}{2}}\right)^2}
\right]\\
&=&\prod_{n=1}^{\infty}
\frac{\left(1+q^{n-\frac{1}{2}}\right)^2}{\left(1-q^n\right)^2}\cdot
\frac{\left(1-q^{3n-2}\right)^2\left(1-q^{3n-1}\right)^2
\left(1-q^{3n}\right)^4}
{\left(1+q^{3n-\frac{1}{2}}\right)^4\left(1+q^{3n-\frac{5}{2}}\right)^4}.
\nonumber
\end{eqnarray}
We can rewrite the right hand side into the following form
\begin{equation}
\prod_{n=1}^{\infty}
\frac{\left(1-q^{6n}\right)^2\left(1-q^{6n-3}\right)^2
\left(1+q^{6n-\frac{3}{2}}\right)^2\left(1+q^{6n-\frac{9}{2}}\right)^2}
{
\left(1+q^{6n-\frac{1}{2}}\right)^2
\left(1+q^{6n-\frac{5}{2}}\right)^2
\left(1+q^{6n-\frac{7}{2}}\right)^2
\left(1+q^{6n-\frac{11}{2}}\right)^2
}.
\end{equation} 
Then dividing the both sides in (\ref{proc})
by $\prod_{n=1}^{\infty}\left(1-q^{6n}\right)^2$,
the first identity in (\ref{d4id}) can be rewritten as
\begin{eqnarray}\label{prod}
&&\prod_{n=1}^{\infty}
(1-q^{6n-1})(1-q^{6n-5})(1+q^{6n-\frac{5}{2}})^2(1+q^{6n-\frac{7}{2}})^2
\nonumber\\
&&+\;q \;\prod_{n=1}^{\infty}
(1-q^{6n-1})(1-q^{6n-5})(1+q^{6n-\frac{1}{2}})^2(1+q^{6n-\frac{11}{2}})^2\\
&=&
\prod_{n=1}^{\infty}
(1-q^{6n-3})^2(1+q^{6n-\frac{3}{2}})^2(1+q^{6n-\frac{9}{2}})^2\nonumber,
\end{eqnarray}
by cancelling the terms in the denominator.  
Using the Jacobi triple product identity
\begin{equation}
\sum_{n=-\infty}^{\infty}q^{n^2}y^n
=\prod_{m=1}^{\infty}(1-q^{2m})(1+yq^{2m-1})(1+y^{-1}q^{2m-1}),
\end{equation}
with $q\to q^3$ and
$y=-q^2,q^{\frac{1}{2}},q^{\frac{5}{2}},-1,q^{\frac{3}{2}}$,
we can rewrite (\ref{prod}) as 
\begin{eqnarray}
&&\left(\sum_{n=-\infty}^{\infty}(-1)^nq^{3n^2+2n}\right)
\left[\left(\sum_{n=-\infty}^{\infty}q^{3n^2+\frac{1}{2}n}
\right)^2+
\;q\;
\left(\sum_{n=-\infty}^{\infty}q^{3n^2+\frac{5}{2}n}
\right)^2
\right]\nonumber\\
&=&
\left(\sum_{n=-\infty}^{\infty}(-1)^nq^{3n^2}\right)
\left(\sum_{n=-\infty}^{\infty}q^{3n^2+\frac{3}{2}n}
\right)^2.
\end{eqnarray}
Then we can drop the unwanted factor $q$ in the second term in the left
hand side.
Using
\begin{eqnarray}
&&\sum_{n=-\infty}^{\infty}q^{k\left(n+\frac{m}{4k}\right)^2}
= \theta_{m,4k}\:(\tau)+\theta_{m+4k,4k}\:(\tau),\nonumber\\
&&\sum_{n=-\infty}^{\infty}(-1)^nq^{k\left(n+\frac{m}{4k}\right)^2}
= \theta_{m,4k}\:(\tau)-\theta_{m+4k,4k}\:(\tau),
\end{eqnarray}
and
$\theta_{m,k}\:(\tau)=\theta_{-m,k}\:(\tau)=\theta_{2k-m,k}\:(\tau)$,
we obtain
\begin{equation}
\left(\theta_{4,12}-\theta_{8,12}\right)
\left[\left(\theta_{1,12}+\theta_{11,12}\right)^2+
\left(\theta_{5,12}+\theta_{7,12}\right)^2\right]
=
\left(\theta_{0,12}-\theta_{12,12}\right)
\left(\theta_{3,12}+\theta_{9,12}\right)^2. \label{d4p}
\end{equation}

At this stage, we use the following properties of the theta function \cite{ta}
\begin{eqnarray}
\theta_{m,4k}\:(\tau)&=&\sum_{\ell=0}^{2k-1}\theta_{2km+16k^2\ell,16k^3}\:
(\tau),\\
\theta_{m,k}\:(\tau)&=&\theta_{2m,4k}\:(\tau)+\theta_{4k-2m,4k}\:(\tau).
\label{for1}
\end{eqnarray}
Also due to the product formula for the theta function
\begin{equation}
\theta_{m,k}\:(\tau)\;\theta_{m',k'}\:(\tau)
=
\sum_{\ell=1}^{k+k'}\;
\theta_{mk'-m'k+2\ell kk',kk'(k+k')}\:(\tau)\;
\theta_{m+m'+2\ell k,k+k'}\:(\tau),
\end{equation}
we can obtain the useful formula 
\begin{equation}
\left(\theta_{m,2k}\pm\theta_{2k-m,2k}\right)\;
\left(\theta_{m',2k}\pm\theta_{2k-m',2k}\right)\:(\tau)
=
\theta_{\frac{m-m'}{2},k}\;\theta_{\frac{m+m'}{2},k}\:(\tau)
\:\pm \:
\theta_{k-\frac{m+m'}{2},k}\;\theta_{k-\frac{m-m'}{2},k}\:(\tau)
.\label{for2}
\end{equation}

We multiply $\left(\theta_{2,12}-\theta_{10,12}\right)$ with
both sides in (\ref{d4p}), and
using the relation (\ref{for2}) in the following
combination
\begin{eqnarray}
&&\left(\theta_{2,12}-\theta_{10,12}\right)
\left(\theta_{4,12}-\theta_{8,12}\right)
=
\left(\theta_{1,6}-\theta_{5,6}\right)\;\theta_{3,6},\nonumber\\
&&\left(\theta_{0,12}-\theta_{12,12}\right)
\left(\theta_{2,12}-\theta_{10,12}\right)
=\theta_{1,6}^2-\theta_{5,6}^2,\nonumber\\
&&\left(\theta_{1,12}+\theta_{11,12}\right)^2=
\theta_{0,6}\;\theta_{1,6}+\theta_{5,6}\;\theta_{6,6},\nonumber\\
&&\left(\theta_{5,12}+\theta_{7,12}\right)^2=
\theta_{0,6}\;\theta_{5,6}+\theta_{1,6}\;\theta_{6,6},\nonumber\\
&&\left(\theta_{3,12}+\theta_{9,12}\right)^2=
\left(\theta_{0,6}+\theta_{6,6}\right)\;\theta_{3,6},\nonumber
\end{eqnarray}
we can prove the (\ref{d4p}), or the original identity in (\ref{d4id}) exactly.

Other type of the identity in (\ref{d4id}), ${\rm Ch}_{2,0}^{NS,(4)}=B_1C_1$
without summation in the left hand side,
can be easily checked only with the infinite product formula (\ref{inf}).

\subsection*{2. $E_6$ case}
\hspace{5mm}
In the same method using (\ref{inf}) as $D_4$ case, 
we can rewrite (\ref{e6id}) into
\begin{eqnarray}
&&\left[\left(\theta_{8,24}-\theta_{16,24}\right)
\left(\theta_{10,24}-\theta_{14,24}\right)\right]
\left[\left(\theta_{5,24}+\theta_{19,24}\right)^2\right]\nonumber\\
&&+\left[\left(\theta_{8,24}-\theta_{16,24}\right)
\left(\theta_{2,24}-\theta_{22,24}\right)\right]
\left[\left(\theta_{11,24}+\theta_{13,24}\right)^2\right]\\
&=&
\left[\left(\theta_{6,24}-\theta_{18,24}\right)
\left(\theta_{4,24}-\theta_{20,24}\right)\right]
\left[\left(\theta_{9,24}+\theta_{15,24}\right)^2\right].\nonumber
\end{eqnarray}
We use (\ref{for2}) for terms in each square bracket,
and expand both sides.
Then we can explicitly prove the equation (\ref{e6id}).
Other identity can be proved along the similar lines.

\subsection*{3. $E_8$ case}
\hspace{5mm}
We prove (\ref{e8id}).
Using (\ref{inf}), we rewrite (\ref{e8id}) into the following form.
\begin{eqnarray}
&&\left(\theta_{4,60}-\theta_{56,60}\right)
\left(\theta_{14,60}-\theta_{46,60}\right)
\left(\theta_{16,60}-\theta_{44,60}\right)
\left(\theta_{26,60}-\theta_{34,60}\right)\nonumber\\
&&\times
\Biggl[
\left(\theta_{28,60}-\theta_{32,60}\right)
\left(\theta_{11,60}+\theta_{49,60}\right)^2
\left(\theta_{19,60}+\theta_{41,60}\right)^2
\left(\left(\theta_{1,60}+\theta_{59,60}\right)^2
+
\left(\theta_{29,60}+\theta_{31,60}\right)^2
\right)\nonumber\\
&&\quad +
\left(\theta_{8,60}-\theta_{52,60}\right)
\left(\theta_{1,60}+\theta_{59,60}\right)^2
\left(\theta_{29,60}+\theta_{31,60}\right)^2
\left(\left(\theta_{11,60}+\theta_{49,60}\right)^2
+
\left(\theta_{19,60}+\theta_{41,60}\right)^2
\right)\Biggr]\nonumber\\
&=&
\left(\theta_{0,60}-\theta_{60,60}\right)
\left(\theta_{10,60}-\theta_{50,60}\right)
\left(\theta_{12,60}-\theta_{48,60}\right)
\left(\theta_{18,60}-\theta_{42,60}\right)
\left(\theta_{20,60}-\theta_{40,60}\right)\nonumber\\
&&\times
\left(\theta_{3,60}+\theta_{57,60}\right)^2
\left(\theta_{15,60}+\theta_{45,60}\right)^2
\left(\theta_{27,60}+\theta_{33,60}\right)^2.
\end{eqnarray}
Then we multiply $\left(\theta_{0,60}-\theta_{60,60}\right)
\left(\theta_{2,60}-\theta_{58,60}\right)
\left(\theta_{22,60}-\theta_{38,60}\right)$
in both sides, and rewrite 
\begin{eqnarray}\label{last}
&&\left[\left(\theta_{0,60}-\theta_{60,60}\right)
\left(\theta_{22,60}-\theta_{38,60}\right)\right]
\left[\left(\theta_{2,60}-\theta_{58,60}\right)
\left(\theta_{4,60}-\theta_{56,60}\right)\right]\nonumber\\
&&\times
\left[\left(\theta_{14,60}-\theta_{46,60}\right)
\left(\theta_{16,60}-\theta_{44,60}\right)\right]
\left[\left(\theta_{26,60}-\theta_{34,60}\right)
\left(\theta_{28,60}-\theta_{32,60}\right)\right]\nonumber\\
&&\times
\left[\left(\theta_{11,60}+\theta_{49,60}\right)^2\right]
\left[\left(\theta_{19,60}+\theta_{41,60}\right)^2\right]
\left(
\left[\left(\theta_{1,60}+\theta_{59,60}\right)^2\right]
+
\left[\left(\theta_{29,60}+\theta_{31,60}\right)^2\right]
\right)\nonumber\\
&&+
\left[\left(\theta_{0,60}-\theta_{60,60}\right)
\left(\theta_{2,60}-\theta_{58,60}\right)\right]
\left[\left(\theta_{16,60}-\theta_{44,60}\right)
\left(\theta_{22,60}-\theta_{38,60}\right)\right]\nonumber\\
&&\times
\left[\left(\theta_{4,60}-\theta_{56,60}\right)
\left(\theta_{26,60}-\theta_{34,60}\right)\right]
\left[\left(\theta_{8,60}-\theta_{52,60}\right)
\left(\theta_{14,60}-\theta_{46,60}\right)\right]\\
&&\times
\left[\left(\theta_{1,60}+\theta_{59,60}\right)^2\right]
\left[\left(\theta_{29,60}+\theta_{31,60}\right)^2\right]
\left(
\left[\left(\theta_{11,60}+\theta_{49,60}\right)^2\right]
+
\left[\left(\theta_{19,60}+\theta_{41,60}\right)^2\right]
\right)\nonumber\\
&=&
\left[\left(\theta_{0,60}-\theta_{60,60}\right)
\left(\theta_{2,60}-\theta_{58,60}\right)\right]
\left[\left(\theta_{0,60}-\theta_{60,60}\right)
\left(\theta_{22,60}-\theta_{38,60}\right)\right]\nonumber\\
&&\times
\left[\left(\theta_{10,60}-\theta_{50,60}\right)
\left(\theta_{12,60}-\theta_{48,60}\right)\right]
\left[\left(\theta_{18,60}-\theta_{42,60}\right)
\left(\theta_{20,60}-\theta_{40,60}\right)\right]\nonumber\\
&&\times
\left[\left(\theta_{3,60}+\theta_{57,60}\right)^2\right]
\left[\left(\theta_{15,60}+\theta_{45,60}\right)^2\right]
\left[\left(\theta_{27,60}+\theta_{33,60}\right)^2\right]\nonumber.
\end{eqnarray}
We use (\ref{for2}) in order to expand each square bracket.
Then by expanding all the terms in a straightforward way and comparing 
both sides, we can check that the equation (\ref{last}) holds.
Thus we have obtained the complete proof of the identity (\ref{e8id}).

\end{document}